\documentclass[12pt,preprint]{aastex}
\usepackage{amsmath}
\usepackage{graphicx}
\usepackage{epsfig}
\usepackage{url}
\usepackage{indentfirst}
\usepackage{color}
\usepackage{fancyvrb}
\usepackage{natbib}

\newcommand{\psr}{PSR~J2021+4026}
\newcommand{\fermi}{{\it Fermi}}
\begin{document}
\VerbatimFootnotes

\title{Repeated state change of variable gamma-ray pulsar, PSR~J2021+4026}
\author{J.~Takata\altaffilmark{1}, H.H.Wang~\altaffilmark{1}, L.C.C.~Lin\altaffilmark{2},  C.-P.~Hu\altaffilmark{3.4}, C.Y.~Hui\altaffilmark{5}, A.K.H.~Kong\altaffilmark{6}, P.H.T.~Tam\altaffilmark{7}, K.L.~Li\altaffilmark{2,6},  K.S.~Cheng\altaffilmark{8}}
  \email{takata@hust.edu.cn}
 \altaffiltext{1}{School of physics, Huazhong University of Science and Technology, Wuhan, 430074, China}
 \altaffiltext{2}{Department of Physics, UNIST, Ulsan 44919, Korea}
\altaffiltext{3}{JSPS International Research Fellow}
 \altaffiltext{4}{Department of Astronomy, Kyoto University, Oiwakecho, Sakyoku, Kyoto 606-8502, Japan}
 \altaffiltext{5}{Department of Astronomy and Space Science, Chungnam National University, Daejeon 305-764,
   Republic of Korea}
 \altaffiltext{6}{Institute of Astronomy and Department of Physics, National
 Tsing Hua University, Hsinchu, Taiwan}
  \altaffiltext{7}{School of Physics and Astronomy, Sun Yat-Sen University, Zhuhai, 519082, China}
  \altaffiltext{8}{Department of physics, The University of Hong Kong, Pokfulam Road, Hong Kong}

\begin{abstract}
 \psr\ is a radio-quiet gamma-ray pulsar  and the first pulsar that shows
  state change of the gamma-ray emission and spin-down rate. The  state change of \psr\ was first observed at 2011
  October, at which the pulsar  changes the state from high gamma-ray flux/low spin-down rate state to low gamma-ray flux/high spin-down rate state. In December 2014, \psr\ recovered the state before the  2011 state change over a timescale  of a few
  months.  We report that  the long term evolution of   the gamma-ray flux and timing behavior  
  suggests that  \psr\ changed the state near  2018 February 1st
    and entered a new low gamma-ray
    flux/high spin-down rate state. At the 2018  state change,
  the averaged flux dropped from  $(1.29\pm 0.01)\times 10^{-6} {\rm cts~cm^{-2}s^{-1}}$ to  $(1.12\pm 0.01)\times 10^{-6} {\rm cts~cm^{-2}s^{-1}}$, which
  is behavior  similar  to the case of the 2011 event.
  The spin-down rate  has increased by $\sim 3\%$ in the new state
  since  the 2018 state change.
  The shapes of  pulse profile and spectrum  in GeV bands  also changed at the 2018 event, and they are consistent with behavior at the  2011 state change.  Our results probably suggest that  \psr\ is  switching
  between different  states with a timescale of several years,  like some radio pulsars (e.g. PSR~B1828-11).
  \psr\ will provide  a unique  opportunity to study the mechanism of the state switching. 
  
\end{abstract}

\keywords{}

\section{Introduction}
\psr\ (3FGL J2021.5+4026) is one of the brightest radio-quiet  gamma-ray pulsars \citep{abdo09} with
a spin period of $P_s\sim 0.265$s and a spin-down power of $L_{sd}\sim 3.4\times 10^{36}{\rm erg~s^{-1}}$.
This pulsar is also known as the  first variable gamma-ray pulsar (\citealt{Allafort2013}, hereafter Al13). \fermi-Large Area Telescope (\fermi-LAT) observed  a sudden gamma-ray flux drop near MJD~55,850 (2011 October) and a change of 
the observed flux ($>$ 100MeV)  from $(8.33\pm 0.08)\times 10^{-10}\rm{erg~cm^{-2}~s^{-1}}$ to $(6.83\pm 0.13)\times 10^{-10}\rm{erg~cm^{-2}~s^{-1}}$  over a timescale of less than a week.
The timing analysis also revealed  that the spin-down rate
 changed from $-(7.8\pm 0.1)\times 10^{-13}\rm{Hz~s^{-1}}$ to $-(8.1\pm 0.1)\times 10^{-13}\rm{Hz~s^{-1}}$. Subsequent
studies \citep{ng16,zh17} found that the gamma-ray flux and the spin-down rate did not show a gradual recovery toward the values  before  2011 October, and  the pulsar stayed at low gamma-ray flux and high spin-down rate state (hereafter LGF/HSD state) for about three years after the  state change. These flux and timing behaviors suggest that the state change of \psr\
  is owing to the change of the structure of the global magnetosphere.  Around 2014 December, \psr\ returned to the high gamma-ray flux/low spin-down rate state (hereafter HGF/LSD state)  over a timescale of a few months.

 \psr\ stayed at the HGF/LSD state  for about three years since 2014 December.  The aperture photometry light curve (time resolution of 30~days) of \psr\  provided by the $\fermi$ science team\footnote{\verb|https://fermi.gsfc.nasa.gov/ssc/data/access/lat/4yr_catalog/ap_lcs/lightcurve_3FGLJ2021.5p4026.png|} clearly  shows that  a gamma-ray flux change occurred around  MJD 58,150 (2018 February 1st) and that \psr\ entered  new low gamma-ray flux state.  Considering that the previous flux change accompanied with a change in the spin-down rate, it is likely that there was another state change in the spin-down rate near MJD~58,150. In the previous study \citep{zh17}, we investigated the evolution of the gamma-ray flux and spin-down rate until MJD~57,700 (2016 November 8th).  In this study, therefore, we  perform timing and spectral analyses for gamma-ray emission of \psr\ with the data after MJD~57,700 in sections~\ref{reduction} and~\ref{result}. We will  discuss the possible mechanism of the state switching  of the \psr\ in section~\ref{disc}.

 \section{$Fermi$ data reduction}
 \label{reduction}
      We use  the \fermi-LAT Pass 8 (P8R2) data in the energy range of 0.1-300 GeV  for the data  analysis. To obtain the
         temporal evolution of the flux and spin-down rate, we select  the data from 2016 November 8 to 2019  March 1 in a $20^{\circ}$ radius region of interest (ROI) centered
 at the 3FGL~J2021.5+4026 position.  The light curve of  \psr\  is obtained at $E > $0.1GeV using the binned likelihood analysis in the \fermi\  Science Tools.
 We select events in the class for the point source or Galactic diffuse analysis (event class=128) and consider the photons collected in the front and back sections of the tracker (evttype =3). We exclude  events with zenith angles larger than $90^{\circ}$ to reduce the Earth's albedo gamma-ray contamination.

 For the spectral analysis, we input spectral  models, which are provided by the \fermi\ Science Support Center\footnote{https\
   ://fermi.gsfc.nasa.gov/ssc/},  for the galactic diffuse emission (gll\_iem\_v06), the isotropic diffuse emission (iso\_P8R2\_SOURCE\_V6\_v06)  and all the 3FGL catalog sources within $20^{\circ}$ of \psr\  to
 account for the spectral contribution. We also include the extended sources, i.e. the Cygnus-X cocoon and Cygnus Loop, near the \psr.   We first fit our data over the entire time range and obtain  spectral parameters
 for the input sources.   In this study, the spectrum of  \psr\
 is modeled with a  power law plus exponential cut-off function of the form
 \begin{equation}
   dN/dE\propto E^{-\Gamma}\exp[-(E/E_c)],
   \label{fitfunc}
 \end{equation}
 where $E$ is the energy of the photon, $\Gamma$ is the photon index and $E_c$ is the cut-off energy. To obtain the best-fitting
 spectral parameters, we perform a binned likelihood analysis using
 the \fermi\ Science Tools \verb|gtlike|. We divide the entire energy range (0.1-300GeV)
 into 12 energy bins with a logarithmic scale, and  best-fit spectral parameters are  obtained with the elimination
  of insignificant sources $(<3\sigma)$. The 1$\sigma$ uncertainty of
 the fitting parameters is provided by \verb|gtlike| tool, and it is calculated by the method discussed in \citep{matt96}. 

 We divide the entire time  range (from 2016 November 8 to  2019  March 1) into 30 day time bins and obtain the long temporal evolution of the flux in $E>0.1$GeV bands.  For each bin,  the contribution of each
 background  source is calculated with the spectral parameters that are obtained from  the entire  data. For \psr, we refit the data by
 the binned likelihood analysis (\verb|gtlike|) and estimate the flux.

 To obtain ephemeris and the pulse profile of \psr\ after 2018 state
 change, we extract the source events within 1$^{\circ}$ radius centered at the target. The time of arrival of all photons  are  corrected to TDB (barycentric dynamical time) with the task \verb|gtbary| supported by {\it Fermi} Science Tools. 

 \section{Results}
 \label{result}
\begin{figure}
  \centering
  \epsscale{1}
  \includegraphics[scale=0.8]{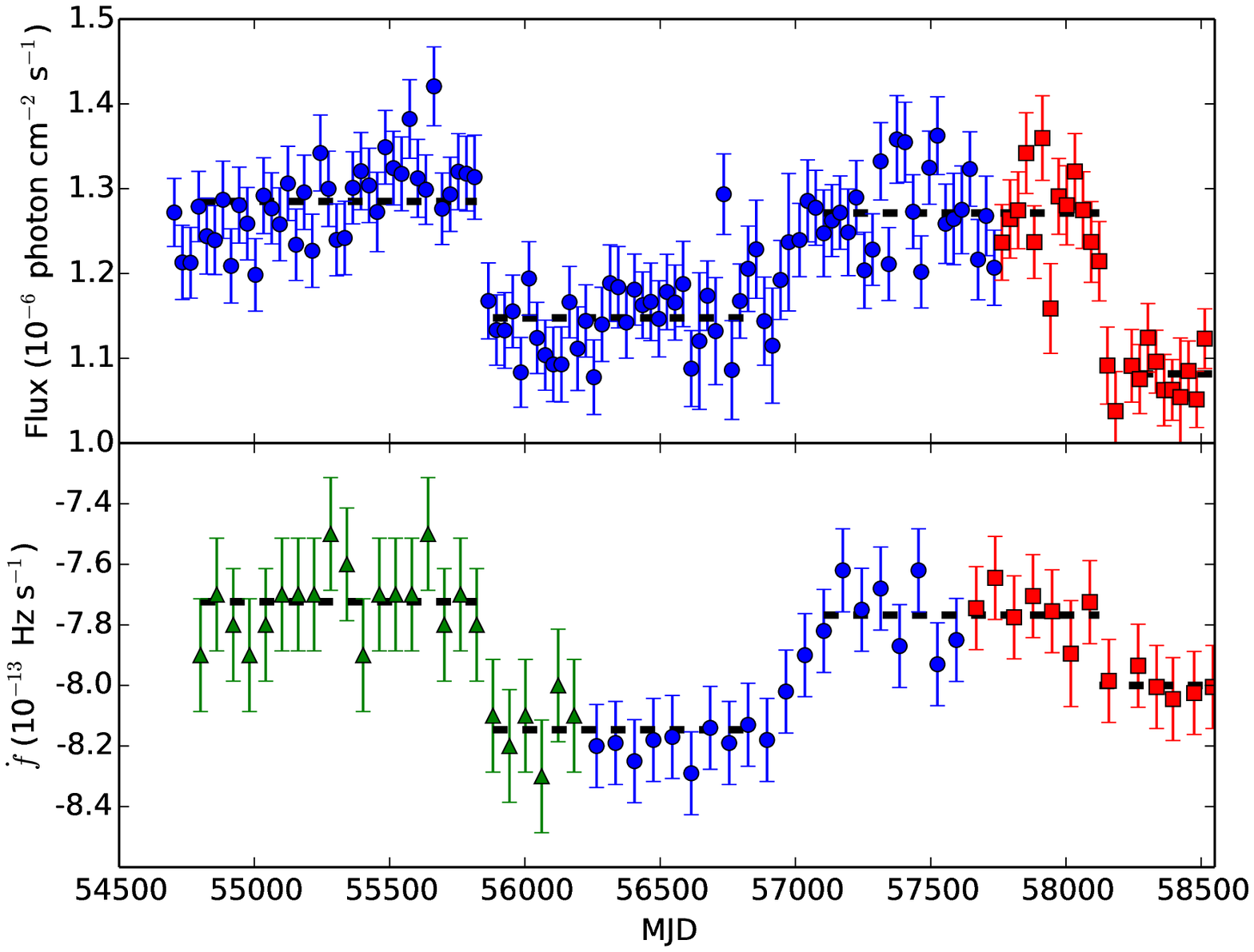}
  \caption{Long term evolutions for $>$0.1GeV flux (upper panel)
    and the spin-down rate $\dot{f}$ (bottom panel).  In the panels, the  triangle (green), circle (blue) and square (red)  symbols
      represent the results obtained by Al13, 
      \cite{zh17}, and this work,  respectively.      The flux (upper panel) is calculated      with  30 day time bin, and the
    spin-down rate (bottom panel) is obtained from  the data with 60 day time bin for Al13  and with 70 day time bin for \cite{zh17}  and this work. 
    The horizontal dashed-line in each panel and each stage denote
    the mean observed flux and spin-down rate. New results show that \psr\ experienced a state change around
    MJD~58,150 (2018 February) again. }
  \label{evolve}
\end{figure}

\begin{figure}
  \centering
  \epsscale{1}
  \includegraphics[scale=0.5]{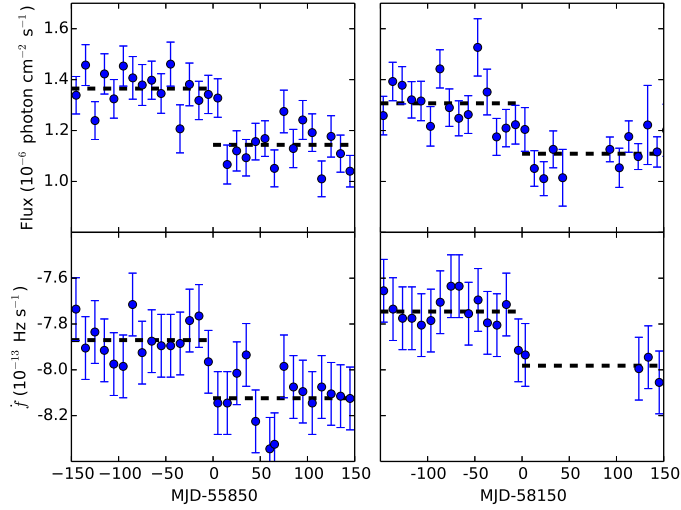}
  \caption{Evolutions for $>0.1$GeV flux (upper panels) and the spin-down rate $\dot{f}$ (bottom panels) around
    the two  state change.  The flux is  calculated with the data of 10~day time bin. In the bottom panel,
      the spin-down rate at each point is obtained with the data of 70 day time bin and the difference between the staring times of
      two neighbor points is 10 days, that is,  two neighbor points share
      the data of 60~days.
    No  $Fermi$-LAT data are available at around MJD~58,200. }
  \label{10day}
\end{figure}

\begin{figure}
  \centering
  \epsscale{1}
  \includegraphics[scale=0.5]{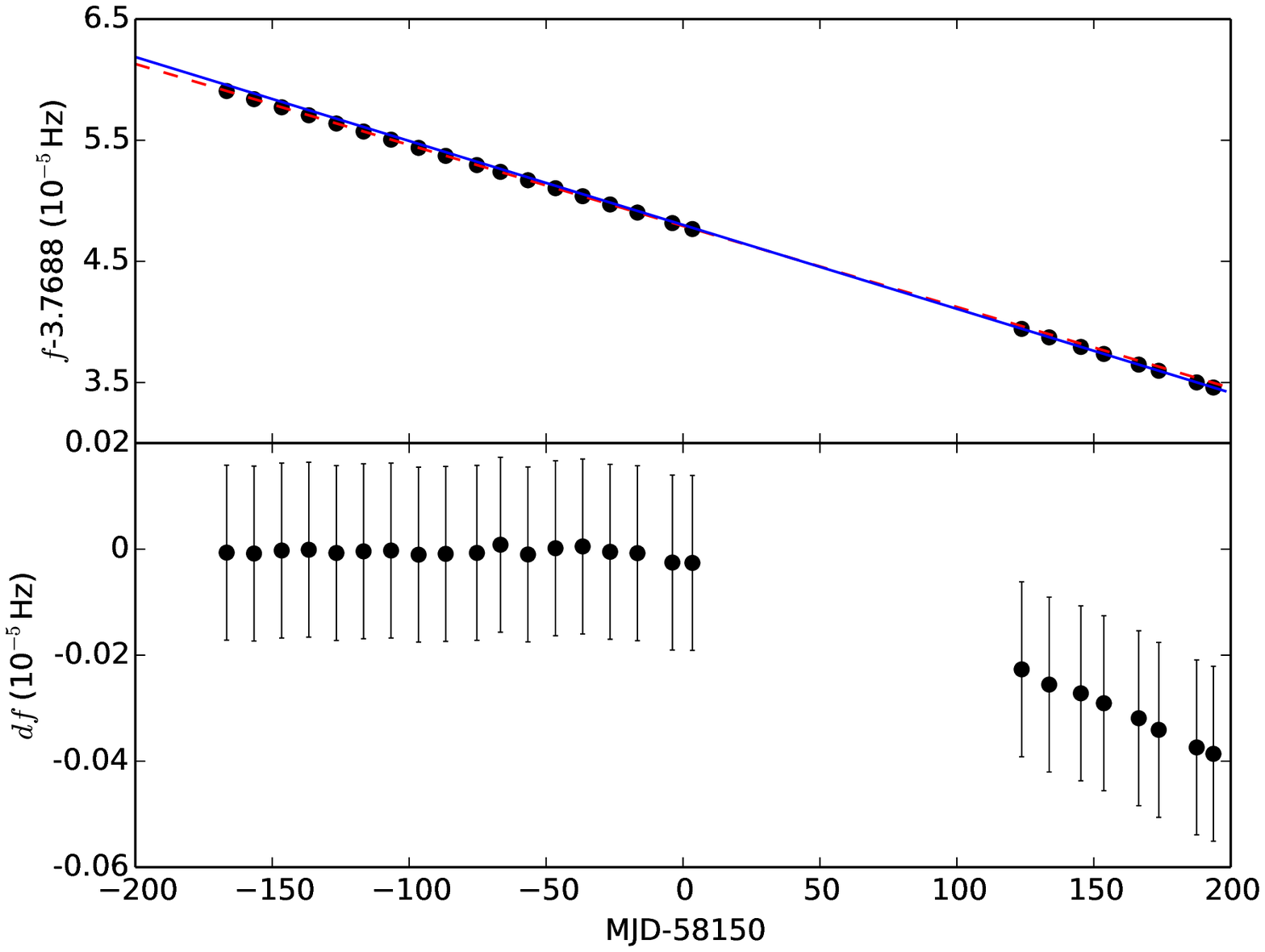}
  \caption{Spin frequency of \psr. Upper : The evolution of the spin frequency around 2018 state change. The dashed (red) and solid (blue) lines
    show the liner functions with a slope, $\dot{f}$, in the middle line (before state change)
    and in the last line (after state change) in Table 1. Bottom : Difference between the observed frequency and predicted frequency from the ephemeris (dashed line in the upper panel) before 2018 state change.  The frequency
    is determined with the data of 70 day time bin, and two neighbor points overlap the 60 day data. The
    uncertainty is determined from the Fourier width, 1/70~days.  No $Fermi$-LAT data  are available around MJD~58,200.}
  \label{fev1}
\end{figure}

\subsection{Evolutions of the gamma-ray flux and spin-down rate}
\label{f1evol}
Top panel of Figure~\ref{evolve} shows the long term evolution for the $>0.1$GeV flux with 30 day time  bins that was used in Al13.
In addition to the previous event near MJD~55,850 reported in Al13,
an obvious flux  change occurred at around MJD~58,100-58,200. The  photon
flux dropped by about $13\pm 1$\% from $F=(1.29\pm 0.01)\times 10^{-6} \rm {cts~cm^{-2}s^{-1}}$ averaged over  MJD~57,105-58,120 to $F=(1.12\pm 0.01)\times 10^{-6} \rm {cts~cm^{-2}s^{-1}}$ averaged over  MJD~58,265-58,500.  The size of the flux drop for the  new event is
  consistent with   $11\pm 1$\%  of the previous event within the error (Table~3).

In the previous flux change, a sudden change of the spin-down rate was observed.
To examine the evolution of the spin-down rate at around the epoch of the new flux
drop, we divide the \fermi-LAT data
after $\sim$MJD~57,700 into a 70 day time  bin,  which can provide enough photons to detect the pulsation \citep{zh17}. During the analysis,
we notice that no \fermi-LAT data for \psr\ is available between MJD~58,190
and 58,240, and we exclude this large observational gap when we bin the \fermi\ data. For each time bin, we describe the frequency
evolution as $f(t)=f(t_0)+\dot{f}(t_0)(t-t_0)$, where $f$,
$\dot{f}$ and $t_0$ represent the spin frequency, the frequency time derivative and reference time, respectively. For reference time ($t_0$), we choose the middle point of each 70 day time  bin.  We perform an $H$-test \citep{DB2010}
to search the best frequency $f(t_0)$ and the time derivative $\dot{f}(t_0)$ in the range of [3.767Hz,3.769Hz]
and  of [$-9\times 10^{-13}$Hz/s, $-7\times 10^{-13}$Hz/s], respectively, within which the best solutions are expected  from the previous analyses (Al13; Zhao et al. 2017).
The bottom panel of Figure~\ref{evolve} shows the evolution of $\dot{f}$, and the uncertainty of each data point  denotes
the Fourier resolution  in the periodicity search.  In each data set, we can clearly identify the pulse
peak in the folded light curve with the local ephemeris.

In Figure~\ref{evolve}, we can see a change of the spin-down rate ($\dot{f}$) around the epoch of the gamma-ray flux drop.
No recovery of the spin-down rate and gamma-ray flux are observed since $\sim$MJD~58,150. This indicates that \psr\ changed
  the state from a HGF/LSD state  to a LGF/HSD state around  MJD~58,150 (2018 February 1st).
  The averaged spin-down rates at the four stages  (dashed lines  in the bottom panel of Figure~\ref{evolve}), are  $\dot{f}=-7.72\pm 0.05\times 10^{-13}\rm{Hz~s^{-1}}$, $-8.15\pm0.04\times 10^{-13}\rm{Hz~s^{-1}}$, $-7.77\pm 0.05\times 10^{-13}\rm{Hz~s^{-1}}$ and   $-8.01\pm 0.07\times 10^{-13}\rm{Hz~s^{-1}}$, 
  which  are calculated from the data points at MJD~54,800-55,820 (HGF/LSD), 55,880-56,825 (LGF/HSD), 57,100-58,120 (HGF/LSD)  and 58,270-58,540 (LGF/HSD), respectively. We find that the spin-down rate in the new HSD state after the 2018 state change  is smaller than that of 2011 state change; the increase of the spin-down rate at 2018 state change,
  $\triangle\dot{f}/\dot{f}=3.1\pm 1.1\%$, is also smaller than the 2011 event  ($5.6\pm 0.9$\%).

Figure~\ref{10day}  is a close-up of  the evolution of the gamma-ray flux with 10 day time bin size (top panels) and the spin-down rate (bottom panels) around  two state change events.
 The spin-down rate  at each point is calculated with the data of 70 day time bin and the starting time  difference between two neighbor points is 10 days; namely, two neighbor points share the data of 60 days.  For the previous event, we can see an abrupt change of  the gamma-ray flux
and the spin-down rate near MJD~55,850 with a timescale less than 10~days (Al13). For the 2018 event, the state change would have occurred at around MJD~58,150 with a timescale similar to the previous event, although  abrupt change of the new event is less clear compared to the previous event.

The state change of \psr\ could be triggered  by the glitch, which is an abrupt change in the spin frequency of the neutron star.
  Figure~\ref{fev1} summarizes  the evolution of the spin frequency around the 2018 state change. The dashed line
  and solid  line in the upper panel of the figure show the liner functions with the slope, $\dot{f}$,  in the middle column (before state change) and in the last column (after state change), respectively,  in Table 1.  The bottom panel shows the difference between the observed spin frequency and
  the predicted one from the ephemeris before the 2018 state change (dashed line in upper panel). We can see a  change of the spin-down rate at around MJD~58,150. With the current uncertainty given by the data, however, we cannot
  confirm  an apparent frequency jump around MJD~58,150. If there is a glitch event, its size, $\triangle f/f$, is  much less than $10^{-7}$. Our result is consistent with the 2011 event (Al13).

\subsection{Pulse profile and phase-averaged spectrum}
The previous studies (Al13; Zhao et al. 2017) fit the gamma-ray pulse profile with two or three Gaussian components, and find that the pulse shape changed at the 2011 event. In particular, the intensity ratio of  Peak~1 and Peak 2 was  measured with  $0.54\pm 0.06$ for the HGF/LSD state and with $0.24\pm 0.03$ for the LGF/HSD state. The third Gaussian component was necessary in the fitting for the pulse profile of the HGF/LSD state before
the 2011  event  and it  was not necessary for the LGF/HSD state.   To examine the pulse profile after the  state change in 2018, we generate the ephemeris between MJD~58,244 (2018, May 6st) and 58,722 (2019, August 27th)  (\citealt{zh17} for the detailed method). Table~1 summarizes the  ephemerides at  different epochs; the second  and third columns present the ephemerides for MJD~55,875-56,943 (previous LGF/HSD state) and 57,062-57,565 (HGF/LSD state) reported in \cite{zh17}, and the fourth column  presents the ephemeris obtained in the current study. As shown  in section~\ref{f1evol},
the spin-down rate after 2018 state change is  smaller than that measured 
after the  2011 event.

\begin{figure}
  \centering
  \epsscale{1}
  \includegraphics[scale=0.5]{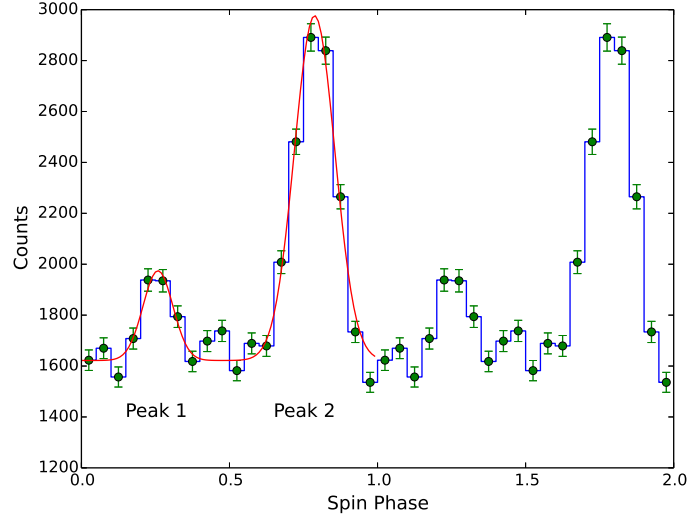}
  \caption{Pulse profile of J2021+4026 in MJD~58,244-58,722 (LGF/HSD state after 2018  state change). The pulse profile is generated with photon energy $>0.1$GeV.  The curved line  show the fitting results with two Gaussian components.}
  \label{light}
  \end{figure}

 Figure~\ref{light} shows a  pulse profile generated with  $>0.1$GeV energy bands after 2018  state change and its fit to the  two Gaussian model  (curved line) is overlaid.  The fitting parameters in the different epochs are summarized in Table~2.
We find that the peak ratio  in the new state after the 2018 event  is $ P1/P2\sim 0.26\pm 0.12$, which is consistent with the previous observation in the LGF/HSD  state after 2011 state change. In the 2011  event, Peak~1 in the pulse profile  changed
the  full width at half maximum (FWHM). In the 2018 event, we also observe the change of FWHM of Peak~1. The
magnitude of the width change is consistent with the 2011 event within the error, as Table~2 shows.

\begin{table*}[t]
   \begin{center}
     \caption[]{Ephemerides of \psr\ derived from LAT data of MJD~55,857--56,943, 57,062--57,565 \citep{zh17} and 58,244--58,722 (this work).      The numbers in parentheses denote 1$\sigma$ errors in the last digit.\\}\label{ephemerides} 
     \begin{tabular}{llll}
       \hline\hline
       \multicolumn{3}{l}{Parameter} \\
       \hline
       Right ascension, $\alpha$\dotfill & \multicolumn{3}{c}{20:21:30.733} \\
       Declination, $\delta$\dotfill &  \multicolumn{3}{c}{+40:26:46.04} \\
       Valid MJD range\dotfill & 55,857--56,943 & 57,062--57,565 & 58,244--58,722\\
       Pulse frequency, $f$ (s$^{-1}$)\dotfill & 3.7689669240(2) & 3.7689112482(6) &  3.7688306913(3) \\
       First derivative, $\dot{f}$ ($10^{-13}$s$^{-2}$)\dotfill & $-$8.1978(1)& $-$7.738(1) & $-$8.0224(4)\\
       second derivative, $\ddot{f}$ ($10^{-22}$s$^{-3}$)\dotfill & $-0.19(2)$ & $3.5(3)$& $-$3.1(1) \\
       Third derivative, $\dddot{f}$ ($10^{-29}$s$^{-4}$)\dotfill & $0.572(7)$ & $-1.7(9)$ & 1.4(2)\\
       Fourth derivative, $10^{-37}f^{(4)}$ (s$^{-4}$)\dotfill & $0.3(1)$ & $-6(9)$ & --\\
       Epoch zero (MJD)\dotfill & 56,400 & 57,200 & 58,400\\
       RMS timing residual ($\mu$s)\dotfill & 2199.970 & 2795.366 & 1808.901 \\
       Time system \dotfill & \multicolumn{3}{c}{TDB (DE405)} \\
       \hline
     \end{tabular}
   \end{center}
 \end{table*}

\begin{table*}[t]
\centering
    \caption[]{Parameters of Gaussian Fitting for Pulse Profiles of $E>0.1$GeV.\\}\label{gauss}
    \begin{tabular}{ccccc}
      \hline\hline
      & \multicolumn{2}{c}{2011 } & \multicolumn{2}{c}{2018} \\
         & HGF/LSD\tablenotemark{c}  & LGF/HSD\tablenotemark{d}  & HGF/LSD\tablenotemark{e} & LGF/HSD \\
      \hline
      Peak 1\tablenotemark{a} & 0.19$\pm$ 0.02 & 0.13$\pm$ 0.02 & 0.195$\pm$0.024 & 0.119$\pm$ 0.027\\
     Peak 2\tablenotemark{a}  & 0.176$\pm 0.007$ & 0.174$\pm$ 0.006   &  0.152$\pm$0.010 & 0.161$\pm$0.013  \\
      Peak 1/Peak 2\tablenotemark{b} & 0.54$\pm$ 0.06 & 0.24$\pm$ 0.03  & 0.494$\pm$0.009 & 0.260$\pm$0.118 \\
      \hline
       \multicolumn{5}{l}{{\footnotesize a: FWHM. b: Ratio of amplitude. c: Al13, three Gaussian.}} \\
       \multicolumn{5}{l}{{\footnotesize d: Al13, two Gaussian. e: Zhao et al. (2017), two Gaussian.}} \\
          
    \end{tabular}
\end{table*}

\begin{figure}
  \centering
  \epsscale{1}
  \includegraphics[scale=0.5]{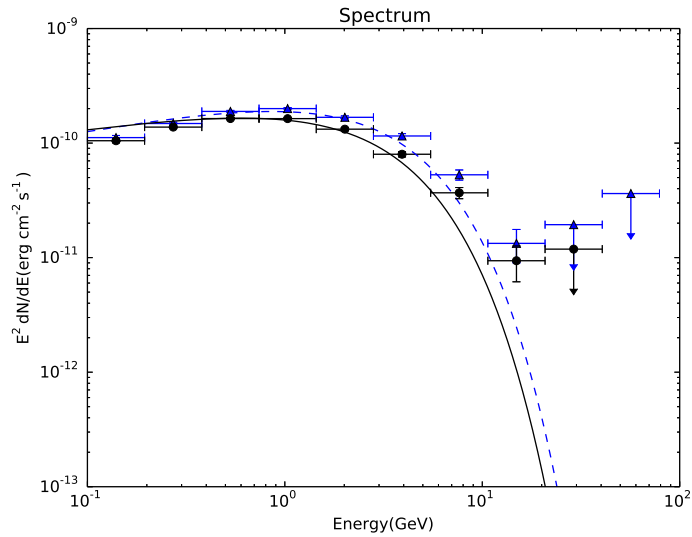}
  \caption{Phase averaged spectra around 2018  state change. Triangle and circle symbols present the spectra in   HGF/LSD state (MJD 57,800-58,100) and
    LGF/HSD state (MJD 58, 250-58,550), respectively.
    The dashed line and solid line show the best fitting function for
    the HGF/LSD state and LFG/HSD state, respectively,  Fitting parameters for each state are presented in Table~3.}
  \label{spec}
  \end{figure}

Figure~\ref{spec} shows the phase-averaged spectra before (MJD 57,800-58,100) and
after (MJD 58,250-58,550) the 2018 state change, and Table~3 summarizes   the  fitting parameters of the different epochs
using a power-law plus exponential cut-off function described by equation
(\ref{fitfunc}). The obvious change in the spectral properties are observed for the 2018  event as well as the 2011  event: the cut-off energy decreased and the photon index increased after the state change.   The similar change of the spectral properties was also observed in the 2011 event, and the observed  flux level and the  cut-off energy between the same states are consistent with each other.  These results therefore suggest that gamma-ray emission of \psr\ repeats the state  change between two different states.
If \psr\  repeats the state change with the current observed time interval, the  next recovery from the current LGF/HSD state to an HGF/LSD state may  occur in 2021 and the next abrupt state change from  the HGF/LSD state to a LGF/HSD state may happen in 2025.

\begin{table}
  \centering
  \caption[]{Parameters of spectral fitting. The parameter in 2011 event are taken from \cite{zh17}.
    \\}
\begin{tabular}{ccc|cc}
  \hline\hline
  & \multicolumn{2}{c|}{2011 } & \multicolumn{2}{c}{2018} \\
  Parameter & HGF/LSD & LGF/HSD & HGF/LSD & LGF/HSD \\
  \hline
  Flux ($10^{-6}\rm{cts~cm^{-2}s^{-1}}$) & 1.29$\pm$0.01 & 1.15$\pm$ 0.01 &
  1.29$\pm$0.01& 1.12$\pm 0.01$ \\
  Cutoff Energy (MeV) &  2755$\pm 74$ & 2477$\pm$ 77   &  2675$\pm$135 & 2447$\pm$ 110 \\
  Photon index & 1.64$\pm 0.011$ & 1.66$\pm$ 0.014  & 1.68$\pm$0.02 & 1.75$\pm$0.02 \\
  \hline
\end{tabular}
\end{table}

\section{Discussion}
\label{disc}
\begin{figure}
  \centering
  \epsscale{1}
  \includegraphics[scale=0.5]{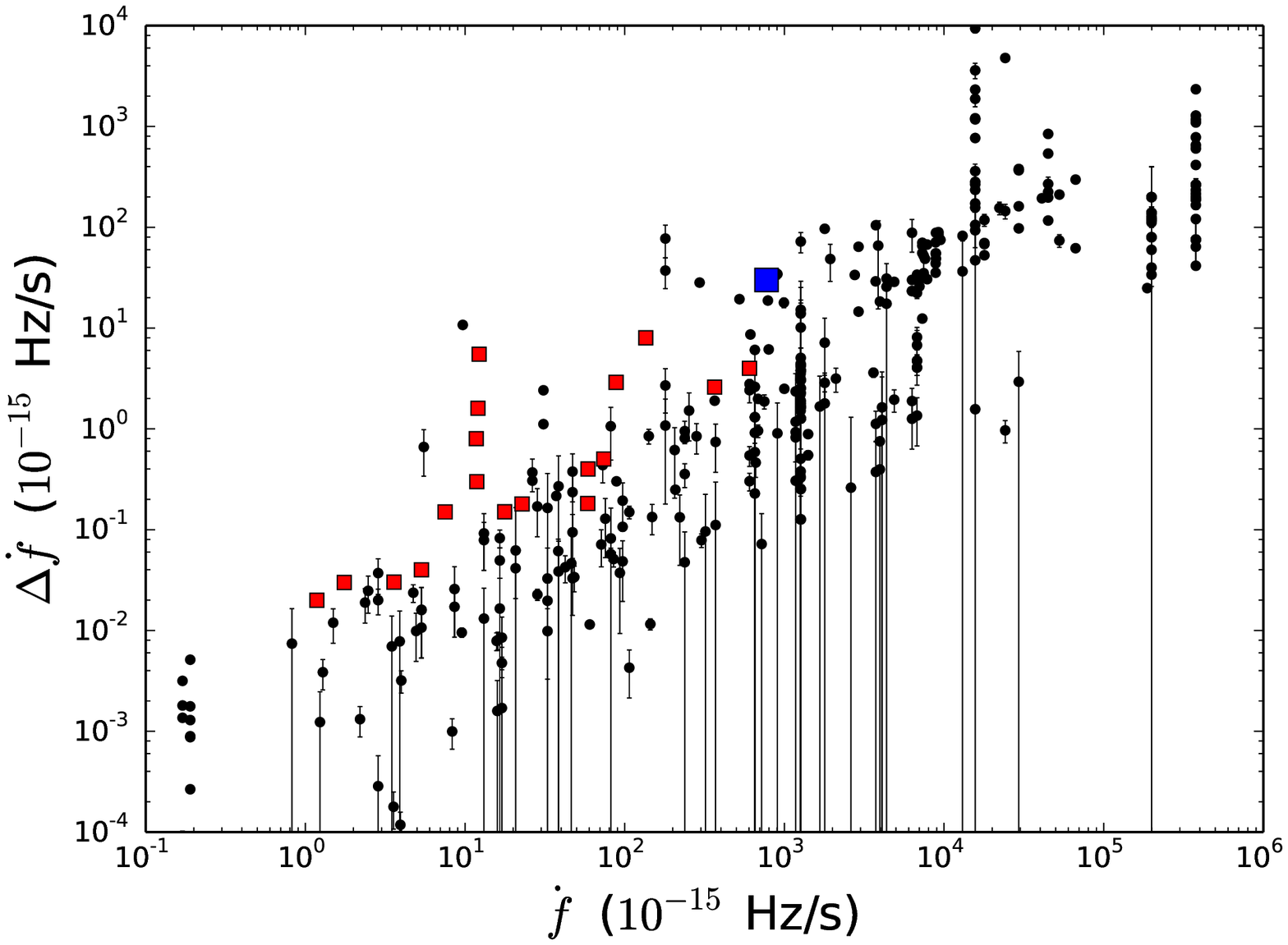}
  \caption{The change of the spin-down rate ($\triangle \dot{f}$) vs the spin-down rate ($\dot{f}$). Blue : \psr.
    Red: Switching pulsar listed in Lyne et al. (2010). Black:
   Glitching pulsars \citep{esp11}. }
  \label{corre}
\end{figure}

The timing and spectral analysis suggest that \psr\ entered a new LGF/HSD  state  at around 2018  February 1st.
Although the spin-down rate of current LGF/HSD state  is slightly smaller than that after 2011~state change,
the gamma-ray emission properties (spectrum and pulse profile)
for the two LGF/HSD states are consistent to each other. Moreover,  the
properties of the emission properties and  spin-down rate between
previous two HGF/LSD states (i) before 2011 October and (ii) between 2014 December and 2018 February  are also consistent to each other.   The long-term
observation of \fermi-LAT suggests  that (i)  \psr\ may repeat an abrupt state change from a HGF/LSD state to a LGF/HSD state
with an time interval of $\tau_{int}\sim 6-7$yrs, (ii) the abrupt state change occurs  over a timescale less than 10~days, and
(iii) \psr\ may stay at an LGF/HSD state for  about  three  years and recover to an HGF/LSD over a timescale of a few months.
Future continuum  monitoring is required to obtain the solid conclusion of the time interval and timescale  of the state change of \psr.  

In this study, we cannot confirm the existence of the glitch at
  the state change. As we will discuss in this section, however, we expect
  that the state change is related to the neutron star's quake,
  which is probably related to the glitch of some pulsars.
  More than 180 glitching pulsars have been confirmed \citep{esp11}\footnote{\verb|http://www.jb.man.ac.uk/pulsar/glitches/gTable.html|}  and about 50 \fermi-LAT pulsars show the glitch event \citep{kerr15}.
However, the state changes of the gamma-ray emissions from other pulsars
have not been confirmed yet. In the radio bands, on the other hand, it has been suggested that various observed phenomena, including emission mode change, nulling, and some timing noise,
are related to the state change of the global magnetosphere. \cite{lyn10} report   timing noise
for 17 radio emitting pulsars, including the glitching pulsars,  and they suggest that those
timing behaviors are  a result of switching (suddenly for some pulsars) between two different  spin-down rates.
They also demonstrate that the change of the spin-down rate of those ``switching pulsars''
is quasiperiodic with a timescale of days to years. Moreover, the timing noise of six pulsars is correlated
to the change of the radio pulse profile, unlike the usual glitch in which no significant
change of the pulse profile is observed. These results suggest that the  structure of global magnetosphere of the state switching pulsars  changes periodically  with time. Although confirmation of the next state change from the current LGF/HSD state to HGF/LSD state is necessary,
\psr\ may also   periodically  switch between two different spin-down states. 

It has been known that the change of the spin-down rate $|\triangle \dot{f}|$ of  the glitching pulsars and  switching  pulsars \citep{lyn10, esp11}  is tightly correlated  with the spin-down rate $|\dot{f}|$ (Figure~\ref{corre}).  This would
  indicate that the change of the spin-down rate of switching pulsars is caused by a process similar to the glitch of some pulsars. 
We can see in Figure~\ref{corre}  that $|\triangle\dot{f}|$ of the switching pulsars tends to have larger  value
than the averaged values of the glitching pulsars, and  the magnitude of $|\triangle\dot{f}|$  of \psr\ is also larger than the values of the glitching pulsars with  a similar spin-down rate.  \psr\ therefore may provide another opportunity to study
the mechanisms of  the state switching. 

Besides the  usual glitch process,  several mechanisms have been proposed to change the spin-down rate of the pulsars.  
For example, \cite{ng16} attribute  the permanent-like increase in the spin-down rate of \psr\  to the change  of the inclination angle by $\triangle \alpha\sim 5^{\circ}$.
The change of the spin-down rate can be also  caused by the change of the global current structure.  The spin-down power may be  described as $L_{sd}\sim IV_{a}$, where $I$ is the current and $V_{a}\sim 2\pi^2\mu/P_s^2c^2$,  with $\mu$ being the magnetic dipole moment of the neutron star,  is the available potential drop of the pulsar. The observed large change in the spin-down rate of \psr\
  may imply the change in the global current rather than the potential drop and
 the current in HSD sate  increases by  $\triangle I/I\sim \triangle \dot{f}/\dot{f}\sim 0.03$.  In the pulsar magnetosphere, the magnitude of the
current would be controlled  by microphysical processes (e.g. pair-creation) of the particle acceleration regions,
where the electric field along the magnetic field line exists. Several possible regions have been suggested for the acceleration/emission region, e.g.
polar cap accelerator \citep{rud75,dau96}, slot gap \citep{ha08, ha15},  outer gap accelerator \citep{ch86,tak16} and current sheet region \citep{spi06}. Since those different acceleration regions will be connected by the global current
flow \citep{shi95}, a state change  of one acceleration region will affect to other acceleration/emission regions and hence to the global magnetosphere structure.
Since it has been thought that the gamma-ray emission
occurs  around the light cylinder, we may expect  that some physical process
around the light cylinder (e.g. magnetic reconnection at the Y-point, magnetosphere  instability) triggers the state change of the pulsar. 
However,  \fermi-LAT has already discovered $>200$  gamma-ray emitting pulsars \citep{fermi19}, but 
only \psr\ shows the switching behavior in both spin-down rate and emission properties (two $Fermi$-LAT pulsars,  PSR J2043+2740 and J0742-2822, are probably switching pulsars,  \citealt{lyn10}). Since \psr\ has typical spin-down parameters  in  the gamma-ray pulsars,  we should expect  more variable
gamma-ray pulsars, if the physical process around the light cylinder could trigger the state switching  of \psr. 

The state switching  of \psr\  is likely  caused by change of the  magnetic field structure of  the polar cap region.
Since the the polar cap region covers  only about $0.05 (P_s/0.1{\rm s})^{-1}$\%  of the stellar surface, the probability of the disturbance of the polar cap region by a glitch process (e.g. crust cracking)  may be low.
This could explain why  switching pulsar, especially switching gamma-ray pulsar is rare. In Ruderman (1991a,b), the crust magnetic flux tube penetrates the type II super-conducting core. The neutron vortex lines in the core
  move as the pulsar spins down, and pull the   magnetic flux tubes in the core with them. If the magnetic flux at the crust
  is immobile, the shear stress will be built up at the crust.  The crust cracking will occur if the shear stress exceeds   the crust elastic stress \citep{ch98},
  \begin{equation}
    B_c\delta B/8\pi>\mu_s\theta_{max}\frac{\ell}{R_{NS}} 
    \end{equation}
  where $B_c$ is the average magnetic field strength in the quantized flux tubes  at the core, $\delta B$ is the magnetic field strength creating the shear stress,
  $\mu_s$ is the shear modulus, $\theta_{max}$ is the  strain before the cracking,  $\ell$ is the size of the crust, and
  $R_{NS}$ is the radius of the neutron star. Hence, the magnetic field strength $\delta B$ just before the cracking
  is estimated to be
  \begin{equation}
    \delta B\sim 3\times 10^{11} {\rm G}\left(\frac{\mu_s}{10^{29}\rm{g~cm^{-3}}}\right)\left(\frac{\theta_{max}}{10^{-3}}\right)
    \left(\frac{B_c}{10^{15}{\rm G}}\right)^{-1}\left(\frac{\ell}{10^{5}{\rm cm}}\right).
    \end{equation}
  After the cracking, the cracked plate will move to reduce the stress.  The displacement ($\delta\ell$)  and the time interval ($\tau_{int}$) 
  between two successive  abrupt state changes may be  related with the stellar radius and characteristic age  ($\tau_{sd}$) of the
  pulsar by  $\tau_{int}/\delta \ell \sim \tau_{sd}/R_{NS}$ (Ruderman 1991a,b). The displacement is rewritten as 
  \begin{equation}
    \delta\ell =10^2{\rm cm} \left(\frac{\tau_{int}}{7{\rm yrs}}\right)\left(\frac{\tau_{sd}}{70{\rm kyrs}}\right)^{-1}.
    \end{equation}
  With the parameter  of PSR~J2021+4026, $\tau_{int}\sim 7$yrs and $\tau_s\sim 70$kyrs, the displacement is estimated to be  $\delta \ell\sim 100$cm.

With  the parameters of PSR J2021+4026, the radius of the polar cap is $R_{pc}\sim 2.8\times 10^{4}$~cm, and the
size of the polar cap accelerator along the magnetic field line will be of the order of $\sim 10^3$cm \citep{rud75}. Hence the
sudden displacement $\ell\sim 10^2$cm of the cracked plate could affect to the polar cap structure.  
With a strong magnetic field at the polar cap region, the  quantum electro dynamical  processes such as the magnetic pair-creation
process and/or the photon splitting process can be realized. The mean free path of such a magnetic process is sensitive to the photon energy, the strength of the magnetic field,  and the angle  between the propagation direction
of the photon and direction of the magnetic field \citep{rud75,ti19}. If the crust cracking  (or other process) would change  the magnetic field structure or field strength in the polar cap region, the process affects to the created current
and hence the spin-down rate.  It is expected  that the change of the current is very limited, that is, $\triangle I/I\ll 1$, since
the magnitude of  current $I$ will be  controlled  between the particle acceleration and the screening of the accelerating field by the pair-creation
process and it will be always  close to  the  Goldreich-Julian rate  \citep{gol69}.

The  quasi-periodic change of the timing residual of the radio pulsars  has been widely argued within the
framework of the precession of the neutron star \citep{sta00, link01, jones12, kerr16}.  In the precession interpretation for a biaxial star, the precession period, $P_p$,
is related to the spin period, $P_s$,  by 
\begin{equation}
 \frac{P_s}{P_p}=\frac{I_0}{I_p}\epsilon ,
  \end{equation}
 where $I_0$ and $I_p$  denote the characteristic moment of inertia,
  which neglects  the small difference between different axes,  of the star and of  precessing body, respectively. The ellipticity $\epsilon$ is  defined by $\epsilon=\triangle I_p/I_0$ with  $\triangle I_p$ being the difference  between the largest and smallest moments of inertia about the principal axes.  With $P_s=0.265$s and  $P_p\sim 6-7$~years that is observed time interval from an HGF/LSD state to a LGF/HSD state, the required ellipticity is of the order of  $\epsilon \sim 10^{-9}$ if $I_p\sim I_0$, and it   is
similar to the values for other switching pulsars. The wobble angle $\theta$ is estimated to  be of the order of
$\theta\sim |\triangle \dot{f}|/|\dot{f}|\sim 2^{\circ}$, which is also similar to the required angle of well studied
precession candidate, PSR B1828-11 \citep{link01}. 

We however argued  that the change from an LSD state to an HSD state of \psr\  happened over a timescale of  $\sim$ 10 days or less, which corresponds to  the phase width
$\delta\Phi\sim 10{\rm days}/6.5{\rm years}\sim 0.004$. Such an abrupt change could be difficult to explain with a simple precession interpretation, since  the variations in the spin-down rate given by the precession
would  be expressed with the fundamental and first harmonics of the Fourier components.
\cite{jones12} argues that an avalanche-like process such as the  pair-creation process discussed above is probably enhanced
at the some precession phase and magnifies the observed spin-down rate. Another possibility is that a glitch of \psr\ excites
new free  precession and about three~years of the LGF/HSD represents  damping timescale of the precession.
The damping timescale is estimated as $\tau_d=400-10^4$ precession period \citep{alp88,jones01}.
With $\tau_d=3$~years of \psr, the precession period is estimated to be $P_p\sim 0.1-2.6$~days. In this scenario,  only a  small part of the star will
be precessed, $I_p<<I_{0}$, if the \psr\ has a  typical ellipticity $\epsilon \sim 10^{-8}-10^{-9}$. Unfortunately, it will be
difficult to confirm the precession with $P_p\sim 1$ days for \psr\  because of the radio quiet  source.

The evolution of the spin-down rate is  also discussed with the twisted  pulsar magnetosphere, which are proposed for the evolutions
of the X-ray luminosity  and the spin-down rate  of the magnetar after the  X-ray outburst \citep{tho00, bel09}.
{In this scenario, the neutron star quake} would twist the closed magnetic field lines and the current flowing along the twisted  magnetic field lines  creates a toroidal current
component. The induced toroidal current can  provide an additional magnetic dipole moment, $\mu$. Since the spin-down rate is proportional to $\mu^2$, the twisted magnetosphere can increase the
spin-down rate. This scenario can explain the evolution of the X-ray luminosity after the X-ray outburst of magnetar XTE~J1810-197 \citep{bel09} and the high-magnetic field radio pulsar PSR J1119-6127 (Wang et al. in preparation). For PSR J1119-6127, the pulsar underwent a large glitch at 2016 July  X-ray outburst \citep{arc16}, and  the spin-down rate rapidly increased by a factor of $\sim 5-10$  in a timescale of a month. Then the pulsar recovered to the pre-burst rate over the following three months \citep{dai18}.
In the twisted pulsar magnetosphere scenario, the timescale that the twisted magnetic field exists is characterized
by the evolution timescale of  '$j$-bundle', and it  for a  canonical pulsar may be  estimated as 
\begin{equation}
  t_t\sim \frac{\phi_0 \sin^2\theta \mu}{cR_{NS}V_e}\sim 7{\rm hours}
  \times  \phi_0\left(\frac{\sin^2\theta}{0.5}\right)
  \left(\frac{\mu}{5\cdot 10^{30}{\rm G~cm^3}}\right)
  \left(\frac{eV_e}{10^{12}{\rm eV}}\right)^{-1},
\end{equation}
where $\phi_0\sim B_t/B$ is the twist angle with $B_t$ being the toroidal field, $\theta$ is the polar angle of the $j$-bundle, $R_{NS}$ is the neutron star radius and $V_e$ is the voltage to maintain the current flowing along the twisted magnetic field lines. For a  canonical pulsar, the potential drop along the magnetic field line
will  develop so that the electrons/positrons emit the GeV photons via the curvature radiation process. The GeV photons will be converted into the electron/positron pairs by the magnetic pair-creation process and the pairs will stop the development of the potential drop along the magnetic field line. Near the stellar surface, the typical magnitude of the potential drop is of the order of $eV_e\sim 10^{12}$eV. With the typical dipole moment, $\mu\sim 5\times 10^{30} {\rm G~cm^3}$,  of canonical pulsars, the timescale for the existence of the twisted magnetic field will be less than one day, which is difficult to explain the observed timescale $\sim 3$ years of high spin-down state of \psr.   It is expected that  if  the abrupt  state change of \psr\ is caused by the twisting of the magnetosphere,  the emission from the heated footprint contributes to the observed X-ray emission in a HSD state. In the previous studies of the X-ray emission from \psr\ \citep{lin13, wa18}, there is no apparent difference  between the X-ray emission properties in the HSD  and LSD states.  It is therefore that  the twisted magnetosphere will  not be the reason of the state change  of \psr.

\section{Summary}
 \psr\ changed the state from a HGF/LSD state to a  LGF/HSD state around MJD~58,150 (2018 February 1st), and
  the state change would occur over a timescale less than 10~days. The properties of the state change are similar to what observed in 2011 October, although  the measured spin-down rate in the new state
  is slightly smaller than that measured after 2011  event.
  The long term evolution of the spin-down rate and gamma-ray emission properties suggest that \psr\  repeats the state change every several years. The future  monitoring is required to obtain the solid conclusion of time interval of the state change of \psr. 

 More than a dozen of switching pulsars have been discovered  in the radio bands. \psr\ will be  the first  switching pulsar discovered in the gamma-ray bands  and provides unique opportunity to understand the mechanisms of the state switching. We argued that the shear stress
building up at the crust will crack the plate that covers the polar cap region.  The displacement of the cracked plate, $\delta \ell\sim 10^2$cm,  will affect to the structure of the polar cap accelerator, whose size along the magnetic field line will be  $\sim10^{3}$cm. It would be also possible that the glitch of the neutron star  excites new precession with a period of 0.1-2.6 days  and
the timescale of LGF/HSD state ($\sim 3$ yrs) would  present the damping timescale of the precession.
If \psr\  repeats the state change with the current observed time interval,  the next recovery from the current LGF/HSD state
  to another HGF/LSD state may occur in 2021 and the next abrupt state change for a HGF/LSD state to a LGF/HSD state may happen in 2025.

  We express our appreciation to an anonymous referee for useful comments and suggestions. J.T. and W.H.H. are  supported by NSFC grants of the Chinese Government under 11573010, 11661161010, U1631103, and U1838102. A.K.H.K. is supported by the Ministry of Science and Technology of the Republic of China (Taiwan) through grants 105- 2119-M-007-028-MY3 and 106-2628-M-007-005.  K.S.C. is supported by GRF grant under 17302315.  C.Y.H. and K.L.L are supported by the National Research Foundation of Korea grant 2016R1A5A1013277. P.-H. T. T. is supported by NSFC through grants 11633007 and 11661161010 and U1731136.


\begin{thebibliography}{}
  \expandafter\ifx\csname natexlab\endcsname\relax\def\natexlab#1{#1}\fi
\bibitem[{Abdo} {et~al.}(2009)]{abdo09}
  {Abdo}, A.~A., {Ackermann}, M., {Ajello}, M., {Atwood}, W.~B., {Axelsson}, M. {et~al.},
  2009a, \apjs, 183, 46

\bibitem[{{Allafort} {et~al.}(2013){Allafort}, {Baldini}, {Ballet},
    {Barbiellini}, {Baring}, {Bastieri}, {Bellazzini}, {Bonamente}, {Bottacini},
    {Brandt}, {Bregeon}, {Bruel}, {Buehler}, {Buson}, {Caliandro}, {Cameron},
    {Caraveo}, {Cecchi}, {Chaves}, {Chekhtman}, {Chiang}, {Chiaro}, {Ciprini},
    {Claus}, {D'Ammando}, {de Palma}, {Digel}, {Di Venere}, {Drell}, {Favuzzi},
    {Ferrara}, {Franckowiak}, {Fusco}, {Gargano}, {Gasparrini}, {Giglietto},
    {Giroletti}, {Glanzman}, {Godfrey}, {Grenier}, {Guiriec}, {Hadasch},
    {Harding}, {Hayashida}, {Hayashi}, {Hays}, {Hewitt}, {Hill}, {Horan}, {Hou},
    {Jogler}, {Johnson}, {Johnson}, {Kerr}, {Kn{\"o}dlseder}, {Kuss}, {Lande},
    {Larsson}, {Latronico}, {Lemoine-Goumard}, {Longo}, {Loparco}, {Lubrano},
    {Malyshev}, {Marelli}, {Mayer}, {Mazziotta}, {Mehault}, {Mizuno}, {Monzani},
    {Morselli}, {Murgia}, {Nemmen}, {Nuss}, {Ohsugi}, {Omodei}, {Orienti},
    {Orlando}, {Paneque}, {Pesce-Rollins}, {Pierbattista}, {Piron}, {Pivato},
    {Porter}, {Rain{\`o}}, {Rando}, {Ray}, {Razzano}, {Reimer}, {Reposeur},
    {Romani}, {Sartori}, {Saz Parkinson}, {Sgr{\`o}}, {Siskind}, {Smith},
    {Spinelli}, {Strong}, {Takahashi}, {Thayer}, {Thompson}, {Tibaldo},
    {Tinivella}, {Torres}, {Tosti}, {Uchiyama}, {Usher}, {Vandenbroucke},
    {Vasileiou}, {Venter}, {Vianello}, {Vitale}, {Winer}, \&
    {Wood}}]{Allafort2013}
  {Allafort}, A., {et~al.} 2013, \apj Letter, 777, L2
\bibitem[Alpar \& Sauls(1988)]{alp88} Alpar, M.~A., \& Sauls, J.~A.\ 1988, \apj, 327, 723
\bibitem[Archibald et al.(2016)]{arc16} Archibald, R.~F., Kaspi, V.~M., Tendulkar, S.~P., \& Scholz, P.\ 2016, \apjl, 829, L21
\bibitem[Beloborodov(2009)]{bel09} Beloborodov, A.~M.\ 2009, \apj, 703, 1044

\bibitem[{Cheng} {et~al.}(1986)]{ch86} Cheng, K.~S., Ho, C., \& Ruderman, M.\ 1986, \apj, 300, 500
\bibitem[Cheng \& Dai(1998)]{ch98} Cheng, K.~S., \& Dai, Z.~G.\ 1998, Physical Review Letters, 80, 18

\bibitem[{{de Jager} \& {B{\"u}sching}(2010)}]{DB2010}
  {de Jager}, O.~C., \& {B{\"u}sching}, I. 2010, \aap, 517, L9
\bibitem[Dai et al.(2018)]{dai18} Dai, S., Johnston, S., Weltevrede, P., et al.\ 2018, \mnras, 480, 3584
\bibitem[{Daugherty and Harding}(1996)]{dau96}
  {Daugherty}, J.~K. and {Harding}, A.~K., 1996, \apj, 458, 278

\bibitem[{Espinoza} {et~al.}(2011)]{esp11}
  {Espinoza}, C.~M., {Lyne}, A.~G., {Stappers}, B.~W. and
  {Kramer}, M., 2011, \mnras, 414, 1689
\bibitem[The Fermi-LAT collaboration(2019)]{fermi19} The Fermi-LAT collaboration 2019, arXiv:1902.10045

\bibitem[{Goldreich and Julian} (1969)]{gol69}
  {Goldreich}, P. and {Julian}, W.~H., 1969, ApJ, 157, 869
\bibitem[{Harding and Kalapotharakos} (2015)]{ha15}
  Harding, A.K.; Kalapotharakos, C., 2015, ApJ, 811,63
\bibitem[{Harding et al.} (2008)]{ha08}
    Harding, A. K.; Stern, J. V.; Dyks, J.; Frackowiak, M., 2008, ApJ, 680, 1378
\bibitem[Jones(2012)]{jones12} Jones, D.~I.\ 2012, \mnras, 420, 2325
\bibitem[Jones \& Andersson(2001)]{jones01} Jones, D.~I., \& Andersson, N.\ 2001, \mnras, 324, 811
\bibitem[Kerr et al.(2016)]{kerr16} Kerr, M., Hobbs, G., Johnston, S., \& Shannon, R.~M.\ 2016, \mnras, 455, 1845
  \bibitem[Kerr et al.(2015)]{kerr15} Kerr, M., Ray, P.~S., Johnston, S., et al.\ 2015, \apj, 814, 128
\bibitem[Lin et al.(2013)]{lin13} Lin, L.~C.~C., Hui, C.~Y., Hu, C.~P., et al.\ 2013, \apjl, 770, L9

\bibitem[Link \& Epstein(2001)]{link01} Link, B., \& Epstein, R.~I.\ 2001, \apj, 556, 392
\bibitem[{Lyne} {et~al.}(2010)]{lyn10}
  {Lyne}, A., {Hobbs}, G., {Kramer}, M., {Stairs}, I. and
  {Stappers}, B., 2010, Science, 329, 408
\bibitem[Mattox et al.(1996)]{matt96} Mattox, J.~R., Bertsch, D.~L., Chiang, J., et al.\ 1996, \apj, 461, 396
\bibitem[{Ng} {et~al.}(2016)]{ng16}
  {Ng}, C.~W., {Takata}, J. and {Cheng}, K.~S., \apj, 2016, 825, 18  

\bibitem[{Ruderman} (1991a)]{ru91a}
  Ruderman, M.\ 1991a, \apj, 366, 261
\bibitem[Ruderman(1991b)]{ru91b} Ruderman, R.\ 1991b, \apj, 382, 576

\bibitem[{Ruderman and Sutherland} (1975)]{rud75}
  {Ruderman}, M.~A. and {Sutherland}, P.~G., 1975, \apj, 196, 51
\bibitem[Stairs et al.(2000)]{sta00} Stairs, I.~H., Lyne, A.~G., \& Shemar, S.~L.\ 2000, \nat, 406, 484

\bibitem[Shibata(1995)]{shi95} Shibata, S.\ 1995, \mnras, 276, 537
\bibitem [{Spitkovsky} (2006)]{spi06}
  {Spitkovsky}, A., 2006, \apj Letter, 648, 51
\bibitem [{Takata} {et~al.}(2016)]{tak16}
  {Takata}, J., {Ng}, C.~W. and {Cheng}, K.~S., 2016, \mnras, 455, 4249
\bibitem[Thompson et al.(2000)]{tho00} Thompson, C., Duncan, R.~C., Woods, P.~M., et al.\ 2000, \apj, 543, 340
\bibitem [{Timokhin and Harding} (2019)]{ti19} Timokhin, A.~N., \& Harding, A.~K.\ 2019, \apj, 871, 12
\bibitem[Wang et al.(2018)]{wa18} Wang, H.~H., Takata, J., Hu, C.-P., et al.\ 2018, \apj, 856, 98
\bibitem[{Zhao} {et~al.}(2017)]{zh17}Zhao, J., Ng, C.~W., Lin, L.~C.~C., et al.\ 2017, \apj, 842, 53

\end{thebibliography}
\end{document}